# An energy optimization method based on mixed-integer model and variational quantum computing algorithm for faster IMPT

Ya-Nan Zhu, Nimita Shinde, Bowen Lin, Hao Gao

*Abstract*- **Intensity-modulated proton therapy (IMPT) offers superior dose conformity with reduced exposure to surrounding healthy tissues compared to conventional photon therapy. Improving IMPT delivery efficiency reduces motion-related uncertainties, enhances plan robustness, and benefits breath-hold techniques by shortening treatment time. Among various factors, energy switching plays a critical role, making energy layer optimization (ELO) essential. This work develops an energy layer optimization method based on mixed integer model and variational quantum computing algorithm to enhance the efficiency of IMPT. The energy layer optimization problem is modeled as a mixed-integer program, where continuous variables optimize the dose distribution and binary variables indicate energy layer selection. To solve it, iterative convex relaxation decouples the dose-volume constraints, followed by the alternating direction method of multipliers (ADMM) to separate mixed-variable optimization and the minimum monitor unit (MMU) constraint. The resulting beam intensity subproblem, subject to MMU, either admits a closed-form solution or is efficiently solvable via conjugate gradient. The binary subproblem is cast as a quadratic unconstrained binary optimization (QUBO) problem, solvable using variational quantum computing algorithms. With nearly the same plan quality, the proposed method noticeable reduces the number of the used energies. For example, compared to conventional IMPT, QC can reduce the number of energy layers from 61 to 35 in HN case, from 56 to 35 in lung case, and from 59 to 32 to abdomen case. The reduced number of energies also results in fewer delivery time, e.g., the delivery time is reduced from 100.6, 232.0, 185.3 seconds to 90.7, 215.4, 154.0 seconds, respectively.**

*Index Terms*- **energy layer optimization, mixed-integer, variation-l quantum computing**

Manuscript received [Month Day, Year]; revised [Month Day, Year]; accepted [Month Day, Year]. Date of publication [Month Day, Year]; date of current version [Month Day, Year]. This research was carried out under Human Subject Assurance Number 00003411 for University of Kansas in accordance with the principles embodied in the Declaration of Helsinki and in accordance with local statutory requirements. Consent was given for publication by the participants of this study. All authors declare that they have no known conflicts of interest in terms of competing financial interests or personal relationships that could have an influence or are relevant to the work reported in this paper. This work was partially supported by the NIH grants R37CA250921, R01CA261964, and a KUCC physicist-scientist recruiting grant. (Corresponding author: Hao Gao.)

Ya-Nan Zhu, Nimita Shinde and Hao Gao are with the Department of Radiation Oncology, University of Kansas Medical Center, Kansas City, KS, USA. (e-mails: [yananzhu6@gmail.com]; nshinde@kumc.edu; [hao.gao.2012@gmail.com]).

Bowen Lin is with Department of Intervention Medicine, the Second Hospital of Shandong University, Jinan, Shandong, China (e-mails: [linbowencore7@outlook.com]).

## I. INTRODUCTION

INTENSITY-modulated proton therapy (IMPT) with pencil beam scanning (PBS) enables highly conformal dose delivery to tumor targets while effectively reducing the exposure of surrounding healthy tissues compared to photon therapy [1]. Despite these advantages, proton therapy is inherently sensitive to uncertainties [2-4]. Reducing the delivery time of IMPT can help mitigate these uncertainties and enhance motion management. A shorter treatment duration minimizes the likelihood of patient or organ motion, decreases the number of required gating cycles [5], and improves overall treatment precision. Additionally, reducing delivery time lowers the risk of motion-induced dose degradation, potentially decreasing reliance on motion mitigation strategies such as rescanning [6]. Furthermore, a shorter delivery time increases the feasibility of patients maintaining breath-holds throughout treatment [7].

IMPT with the PBS technique delivers dose to the tumor in a spot-by-spot manner within each energy layer, then progressing layer-by-layer. Each individual spot is defined by its lateral position, which is controlled by magnetically steering the proton beam, while the depth position is determined by adjusting the proton beam energy. Unlike lateral scanning, which is achieved through fast magnetic beam deflection and typically completes within a few milliseconds, adjusting the beam energy to reach different depths requires modifying the accelerator or degrader settings, which can take seconds, and it consists of the major part of delivery time. It is potential to reduce the delivery time by decreasing the number of energies. Although increasing energy spacing reduces the number of layers, it may neglect critical energies, leading to degraded plan quality [8,9]. Therefore, the energy reduction should be carefully modeled, and the number of energies should be optimized.

A variety of energy layer optimization methods have been proposed in the literature. For instance, Hillbrand et al. [8] introduced a heuristic guideline for energy layer selection in proton therapy, derived from treatment planning studies conducted on a homogeneous phantom. Their approach offers practical insights into optimizing energy layer selection to enhance treatment efficiency and precision. Van De Water et al. [9,10] formulated an optimization strategy that minimizes the logarithm of the total spot weight per layer, subsequently eliminating low-weighted energy layers. Cao et al. [11] approached the problem using a relaxed mixed integer programming framework, implementing an iterative procedure to sequentially remove energy layers. Jensen et al. [12] introduced a root mean squared (RMS) regularization technique, systematically excluding the energy layer with the lowest RMS value to improve treatment plan efficiency. Lin et al. [13] developed MMSEL, an optimization framework incorporating group sparsity and minimum monitor-unit (MMU) constraints to enforce energy layer reduction while maintaining plan quality. Lin et al. [14] proposed the CARD method, which applies a cardinality-sparsity constraint to optimize energy selection. Wang [15] followed the same modeling framework as [14] and employing block orthogonal matching pursuit for a greedy optimization of energy layers.

In this work, energy layer optimization is formulated as a mixed-integer programming model, where continuous variables govern the dose distribution, and binary variables encode the selection of energy layers. To efficiently solve this problem, we employ the alternating direction method of multipliers (ADMM) to decouple the optimization of continuous and binary variables. The resulting subproblems involving continuous variables either admit closed-form solutions or can be efficiently solved using the conjugate gradient method. The subproblem involving the binary variables is formulated as a quadratic unconstrained binary optimization (QUBO) problem, which is then addressed using a variational quantum computing algorithm.

## II. Method

### A. Mathematical Model

The mathematical model of inverse planning of radiation therapy usually takes the following form

$$\min_{x \in \mathbb{R}^n} f(d, \Omega)$$

$$s.t. \begin{cases} d = Dx \\ x(j) \in \{0\} \cup [G_{\min}, +\infty), j \le n \end{cases} \quad (1)$$

where $D$ is the dose influence matrix, and $x$ is the beam intensity. Each entry $x(j)$ of variable $x$ satisfies the minimum-monitor unit constraint (MMU) [16-18] with positive MMU threshold $G_{min}$. $f(d,\Omega)$ is dose-volume histogram (DVH) [19,20] plan objective, penalizing the deviation between the prescribed dose and optimized dose $d$ for activated voxels (indexed by $\Omega$, see supplementary for details). Using the quadratic objective function, the $f(d,\Omega)$ takes the following form.

$$f(d, \Omega) = \sum_{i=1}^{N_{L^2}} \frac{\omega_{1,i}}{|\Omega_{1,i}|} \left\| d_{\Omega_{1,i}} - d_{1,i} \right\|^2 +$$

$$\sum_{i=1}^{N_{DVH-\max}} \frac{\omega_{2,i}}{|\Omega_{2,i}|} \left\| d_{\Omega_{2,i}} - d_{2,i} \right\|^2 + \quad (2)$$

$$\sum_{i=1}^{N_{DVH-\min}} \frac{\omega_{3,i}}{|\Omega_{3,i}|} \left\| d_{\Omega_{3,i}} - d_{3,i} \right\|^2$$

where $|\cdot|$ denotes the cardinality of a set. The first term represents an L2-type objective, which minimizes the mean squared deviations between the optimized dose and the prescribed dose across the considered structures. The second term corresponds to the DVH-max constraint objective, which imposes an upper bound on the proportion of overdosed voxels within the region of interest (ROI). The third term represents the DVH-min constraint objective, which enforces a lower bound on the proportion of underdosed voxels within the target. $\omega_{1,i}$, $\omega_{2,i}$ and $\omega_{3,i}$ are positive objective weights.

Proton therapy utilizing the PBS technique delivers dose by scanning the target spot by spot, and then layer-by-layer. Since energy switching time accounts for the majority of the total delivery time, one strategy to improve delivery efficiency while maintaining plan quality is to reduce the number of energy layers used. For given dose influence matrix D, based on the principles of pencil beam scanning, we can decompose it group-wisely by column as the follows

$$D = [D_1, D_2, \cdots, D_N] \quad (3)$$

where the submatrix $D_i$ denotes the dose influence matrix for the $i^{th}$ energy with its number of columns indicating the number of spots at that energy. The total number of energies is denoted by $N$. Analogously, the beam intensity $x$ can be decomposed as the follows

$$x = [x_1, x_2, \cdots, x_N]^T \quad (4)$$

where each subvector $x_i$ of $x$ represents the beam intensity of the spots corresponding to the specific energy layer. For a given energy layer, it is considered active if the corresponding beam intensity is nonzero. To explicitly model the selection of energy layers, we introduce an additional binary variable to indicate whether a specific energy layer is utilized, i.e., $s$ is the binary variable such that

$$s(i) = \begin{cases} 1, \text{ if energy layer } i \text{ is used} \\ 0, \text{ otherwise} \end{cases}, i = 1, 2, \cdots, N$$

Then, (1) can be reformulated as

$$\min_{x \in \mathbb{R}^n, s \in \{0,1\}^N} f(d, \Omega)$$

$$s.t. \begin{cases} d = \sum_{i=1}^{N} D_i x_i \cdot s(i) \\ x(j) \in \{0\} \cup [G_{\min}, +\infty), j \le n \end{cases} \quad (5)$$

Furthermore, by introducing the binary variable $s$, one can conveniently restrict the number of active energies by the following equation

$$\sum_{i=1}^{N} s(i) = N_E \quad (6)$$

where the integer $N_E$ is the predefined number of active energy layers.

By combining equations (5) and (6), we obtain the following mixed-integer model for energy layer optimization:

$$\min_{x \in \mathbb{R}^n, s \in \{0,1\}^N} f(d, \Omega)$$

$$s.t. \begin{cases} d = \sum_{i=1}^N D_i x_i \cdot s(i) \\ \sum_{i=1}^N s(i) = N_E \\ x(j) \in \{0\} \cup [G_{\min}, +\infty), \ j \le n \end{cases} \quad (7)$$

The model (7) aims to find the proper beam intensity $x$ to achieve desired the plan quality while ensuring that the number of utilized energy layers remains $N_E$.

### B. Optimization algorithm

The (7) is no easy to solve as it involves non-convex DVH objective and MMU constraint. Also, the optimization variable is mix of continuous variable $x$ and discrete of variable $s$. As in [21-25], we first use iterative convex relaxation to decouple nonconvex DVH objective. Since the active index $\Omega$ depends on dose vector $d$, thus depend on $x$, follow the notation and method in [21-25], in each iteration $\Omega$ is updated by

$$\Omega^{k+1} = H(x^k) \quad (8)$$

Readers may refer to the supplementary section A for more details of $H$.

With $\Omega_{k+1}$ fixed, the (7) can be rewritten as

$$\min_{x \in \mathbb{R}^n, s \in \{0,1\}^N} \left\| \sum_{i=1}^N A_i(\Omega^{k+1}) x_i \cdot s(i) - b \right\|_2^2$$

$$s.t. \begin{cases} \sum_{i=1}^N s(i) = N_E \\ x(j) \in \{0\} \cup [G_{\min}, +\infty), \ j \le n \end{cases} \quad (9)$$

where we rewrite the plan objective into concise form by merging the objective weight and cardinality of $\Omega$ in (2) into the normed square and taking into account the updated active index $\Omega_{k+1}$.

For (9), one can decouple the MMU constraint and optimization of variable $x$, $s$ using ADMM [26-30].

First, reformulate (9) by introducing auxiliary variable $z$ for $x$, one gets

$$\min_{x \in \mathbb{R}^n, s \in \{0,1\}^N} \left\| \sum_{i=1}^N A_i(\Omega^{k+1}) x_i \cdot s(i) - b \right\|_2^2$$

$$s.t. \begin{cases} \sum_{i=1}^N s(i) = N_E \\ x = z \\ z(j) \in \{0\} \cup [G_{\min}, +\infty), \ j \le n \end{cases} \quad (10)$$

The augmented Lagrangian $L(x,s,z,\lambda_1,\lambda_2)$ for (10) is written as

$$L(x, s, z, \lambda_1, \lambda_2) = \left\| \sum_{i=1}^N A_i(\Omega^{k+1}) x_i \cdot s(i) - b \right\|_2^2 +$$

$$\mu_1 \|x - z + \lambda_1\|_2^2 + \mu_2 \left\| \sum_{i=1}^N s(i) - N_E + \lambda_2 \right\|_2^2 \quad (11)$$

Where the positive constant $\mu_1, \mu_2$ are weight of the augmented term. ADMM solve (10) by optimizing $L(x,s,z,\lambda_1,\lambda_2)$ in (11) alternatively.

#### 1. Optimize $x$

$$x^{k+1} = \arg\min_x L(x, s^k, z^k, \lambda_1^k, \lambda_2^k)$$

$$= \arg\min_x \left\| \sum_{i=1}^N A_i(\Omega^{k+1}) x_i \cdot s(i) - b \right\|_2^2 + \mu_1 \|x - z^k + \lambda_1^k\|_2^2 \quad (12)$$

$$= \arg\min_x \left\| \tilde{A}^{k+1} x - b \right\|_2^2 + \mu_1 \|x - z^k + \lambda_1^k\|_2^2$$

where

$$\tilde{A}^{k+1} = \begin{bmatrix} A_1(\Omega^{k+1}) \circ s(1) & \cdots & A_N(\Omega^{k+1}) \circ s(N) \end{bmatrix} \quad (13)$$

and $A_i(\Omega^{k+1}) \circ s(i)$ $(i=1,2,...,N)$ denote a matrix of the same size as $A_i(\Omega^{k+1})$, with each entry scaled by $s(i)$.

By the first-order optimality condition of (12), the $x^{k+1}$ is the solution of the following linear system equations

$$(\tilde{A}^{k+1})^T (\tilde{A}^{k+1} x^{k+1} - b) + \mu_1 (x^{k+1} - z^k + \lambda_1^k) = 0$$

$$\Leftrightarrow \left[ (\tilde{A}^{k+1})^T \tilde{A}^{k+1} + \mu_1 I \right] x^{k+1} = (\tilde{A}^{k+1})^T b + \mu_1 (z^k - \lambda_1^k) \quad (14)$$

Observing the system matrix is positive definite, it can be solved iteratively by conjugate gradient method [31].

#### 2. Optimize $s$

The optimization of $s$ takes the following form

$$s^{k+1} = \arg\min_{s \in \{0,1\}^N} L(x^{k+1}, s, z^k, \lambda_1^k, \lambda_2^k)$$

$$= \arg\min_{s \in \{0,1\}^N} \left\| \sum_{i=1}^N A_i(\Omega^{k+1}) x_i^{k+1} \cdot s_i - b \right\|_2^2 + \mu_2 \left\| \sum_{i=1}^N s_i - N_E + \lambda_2^k \right\|_2^2$$

$$= \arg\min_{s \in \{0,1\}^N} \left\| B^{k+1} s - b \right\|_2^2 + \mu_2 \left\| 1^T s - N_E + \lambda_2^k \right\|_2^2$$

$$= \arg\min_{s \in \{0,1\}^N} s^T \left[ (B^{k+1})^T B^{k+1} + \mu_2 11^T \right] s - 2 \left[ b^T B^{k+1} + \mu_2 (N_E - \lambda_2^k) 1^T \right] s \quad (15)$$

where

$$B^{k+1} = \begin{bmatrix} A_1(\Omega^{k+1}) x_1^{k+1} & \cdots & A_N(\Omega^{k+1}) x_N^{k+1} \end{bmatrix} \quad (16)$$

and $I$ is a vector of all ones. The (15) is a classical quadratic unconstraint binary optimization (the quadratic refer to the objective function) problem which can be solved by variational quantum computing (VQA) algorithms. In this approach, the binary variables are mapped to quantum spin states, and the objective function is encoded as a cost Hamiltonian whose ground state corresponds to the optimal solution. A parameterized quantum circuit alternates between applying problem-specific and mixer Hamiltonians, and a classical optimizer is used to iteratively update the circuit parameters to minimize the expected cost. For a detailed description of VQA, we refer readers to the reference [32].

#### 3. Optimize $z$

The $z$-subproblem has analytical solution as follows

$$z^{k+1} = \begin{cases} \arg\min_z L(x^{k+1}, s^{k+1}, z, \lambda_1^k, \lambda_2^k) \\ s.t.\ z(j) \in \{0\} \cup [G_{\min}, +\infty), j \le n \end{cases}$$

$$= \begin{cases} \arg\min_z \left\| x^{k+1} - z + \lambda_1^k \right\|_2^2 \\ s.t.\ z(j) \in \{0\} \cup [G_{\min}, +\infty), j \le n \end{cases}$$

$$\Rightarrow z^{k+1}(j) = \begin{cases} \max(G_{\min}, (x^{k+1}(j) + \lambda_1^k(j))), (x^{k+1}(j) + \lambda_1^k(j)) \ge \dfrac{G_{\min}}{2} \\ 0, otherwise \end{cases}$$

(17)

## 4. Dual variable $\lambda_1, \lambda_2$

$$\lambda_1^k = \lambda_1^k + x^{k+1} - z^{k+1}$$

$$\lambda_2^k = \lambda_2^k + \sum_{i=1}^N s_i^{k+1} - N_E$$

(18)

### C. Practical algorithm

In former subsection, we give the algorithm that solve the model (7) with the fixed energy layer $N_E$. In practice, the optimal $N_E$ is case-by-case and should be selected properly. We give a practical workflow algorithm for selecting $N_E$ in the Fig. 1.

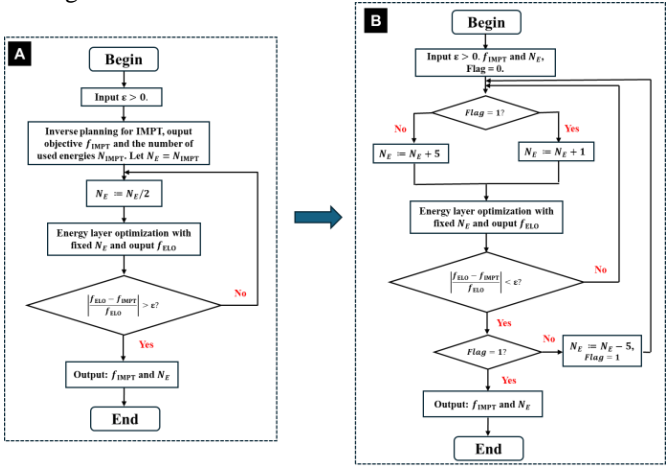

**Fig. 1.** Workflow of practical energy layer optimization method.

Same as [11], the selection of $N_E$ is determined based on the relative error in the plan objective value between IMPT and the ELO method. Initially, we employ a bisection approach to determine the lower bound of $N_E$. Starting with the energy layers used in IMPT, we iteratively reduce $N_E$ by half until the relative error exceeds a predefined threshold $\varepsilon$ (Fig. 1A). Then, we adaptively increase $N_E$ from this lower bound until the relative error fall below $\varepsilon$ (Fig. 1B). To be specific, observing the lower bound obtained from Fig. 1A may be much smaller than the optimal solution, we increment $N_E$ by 5 in each iteration efficiently approximate the optimal $N_E$. Once the relative error criterion is met, we identify the interval in which the optimal $N_E$ lies and further fine-tune $N_E$ to precisely determine the optimal number of energy layers.

### C. Materials

Three clinical cases are presented, including head-and-neck (HN) (2Gy×10; TABLE 1 and Fig. 2), lung (2Gy×30; TABLE 2 and Fig. 2) and abdomen (2Gy×10; TABLE 3 and Fig. 3) cases. Dose influence matrices were generated via matRad [33] with 3 mm lateral spot spacing and 5 mm longitudinal spot spacing on 3 mm$^3$ dose grid. Robust optimization was considered with 5mm (3mm for HN) setup uncertainty and 3.5% range uncertainty which results in 9 scenarios (i.e., $N_s = 9$). Clinically verified beam angles (45º, 130º, 225º, 315º), (0º, 60º, 120º), and (60º, 150º, 240º, 330º) were employed for the HN, lung, and abdomen.

Clinically-used DVH plan objectives were utilized, and all plans were normalized to D$_{95\%}$ = 100% in CTV. The conformity index (CI) was evaluated, defined as V$_{100,CTV}^2$/(V$_{CTV}$ × V$_{100}$). (V$_{100,CTV}$: CTV volume receiving at least 100% of prescription dose; V$_{CTV}$: CTV volume; V$_{100}$: total body volume receiving at least 100% of prescription dose; ideally CI = 1). The dose quantities are in percentage with respect to the prescription dose; the values of objective value and CI are unitless. The delivery time was calculated per fractionation, including the energy switching time (ELST), spots spill time (SSPT) and the spots switching time (SSWT). The energy switching down time is modeled as 0.7 second and switching up is modeled as 5.5 second, ELST is calculated as the total time of switching up and down. The SSPT is approximated as $\|x\|_1/\gamma$, where $\|\cdot\|_1$ is the sum of spot weight and $\gamma$=2.6×10$^{11}$ protons/mins. The calculation of the SSWT is based on a data fitting model, which incorporates line scan switching in the y-axis direction, spot switching along each line in the x-axis direction, and a magnetic preparation time of 1 ms [34,35].

## III. RESULTS

In this section, we evaluate the performance of the proposed method (QC). In subsections A-C, we analyze the impact of reducing the number of energy layers ($N_E$) on plan quality and delivery time for HN, lung, and abdomen cases. In subsection $D$, we perform robust optimization across different cases to assess the method's robustness. Subsection $E$ presents a comparison of QC with state-of-the-art method.

### A. HN

For the HN case, the total number of energy layers is 66. We consider fixed $N_E$ = 20, 30, and 40 to evaluate the performance of QC. As shown in TABLE 1, QC with 30 energy layers achieves a comparable plan quality to IMPT with 60 energies. For example, in terms of target coverage, the max dose and CI are 116.4, 0.76 in IMPT and they are 116.8, 0.74 in QC. The decreased energy layers also do not substantially degrade the OAR sparing, e.g., the mean brainstem and body dose are 4.34 and 4.40 in IMPT while they are 4.64 to 4.41 in QC. However, the decreased energy layers in QC leads to better delivery efficiency, e.g., decreases the delivery time from 98.6 to 79.3 seconds. The configuration with 20 and 10 energy layers achieves the shorter delivery time, reducing the delivery time from 98.6 seconds to 76.9 and 56.7 seconds. However, this comes at the expense of the worse target coverage and OAR sparing as evidenced in TABLE 1. The dose distribution and DVH plots in Fig. 1 further substantiates the results.

TABLE 1. Parameters for HN. The plan-quality dosimetric quantities from top to bottom: number of used energies (Energy), optimization objective value $f$, conformity index (CI), max target dose $D_{max}$, mean brainstem dose ($D_{mean\text{-}brainstem}$), mean body dose ($D_{mean\text{-}body}$), energy layer switching time (ELST), spots spill time (SSPT), spots switching time (SSWT), total delivery time (Total). The unit of time is in second.

| Paras \ Model | IMPT | QC-10 | QC-20 | QC-30 |
|---|---|---|---|---|
| Energy | 60 | 10 | 20 | 30 |
| $f$ | 8.00 | 65.6 | 12.8 | 9.70 |
| CI | 0.76 | 0.57 | 0.69 | 0.74 |
| $D_{max}$ (%) | 116.4 | 129.4 | 117.5 | 116.8 |
| $D_{mean\text{-}brainstem}$ (%) | 4.34 | 8.19 | 5.15 | 4.64 |
| $D_{mean\text{-}body}$ (%) | 4.40 | 5.01 | 4.55 | 4.41 |
| ELST(s) | 46.1 | 49.0 | 44.2 | 44.8 |
| SSPT(s) | 57.8 | 6.30 | 29.8 | 36.8 |
| SSWT(s) | 6.24 | 1.45 | 2.90 | 4.30 |
| Total | 110.1 | 56.7 | 76.9 | 85.9 |

### B. Lung

For the lung case, the total number of energy layers is 61. We evaluate QC using 20, 30, and 40 energy layers, respectively. As shown in TABLE 2, QC with 40 energy layers achieves a comparable plan quality to IMPT with 55 energies, which is shown by nearly the same max target dose and CI. Also, QC achieved a little better OAR sparing than IMPT. When the 30 energy layers are used, with a little degraded target coverage, QC get much better OAR sparing for the heart and spin cord. For example, the mean heart and spin cord dose decreased from 1.72, 3.52 to 0.98, 1.60. When the number of energies is further reduced to 20, although the QC obtained the best OAR sparing, it also gets worst target coverage. Since the reduction in energy layers is less significant than in the HN case, the decrease in delivery time is not that substantial as HN, decreasing from 223.8 to 211.9, 201.8 and 191.1 for the case 40, 30 and 20 energy layers cases. The dose distribution and DVH plots in Fig. 2 further supports these results.

TABLE 2. Parameters for lung. The plan-quality dosimetric quantities from top to bottom: number of used energies, optimization objective value $f$, conformity index (CI), max target dose $D_{max}$, mean lung dose ($D_{mean\text{-}lung}$), mean heart dose ($D_{mean\text{-}heart}$), mean esophagus dose ($D_{mean\text{-}eso}$), mean spin cord dose ($D_{mean\text{-}cord}$), mean body dose ($D_{mean\text{-}body}$), energy layer switching time (ELST), spots spill time (SSPT), spots switching time (SSWT), total delivery time (Total). The unit of time is in second.

| Paras \ Model | IMPT | QC-20 | QC-30 | QC-40 |
|---|---|---|---|---|
| Energy | 55 | 20 | 30 | 40 |
| $f$ | 3.46 | 5.97 | 4.00 | 3.59 |
| CI | 0.89 | 0.80 | 0.87 | 0.89 |
| $D_{max}$ (%) | 116.0 | 122.1 | 117.3 | 116.2 |
| $D_{mean\text{-}lung}$ (%) | 6.98 | 6.74 | 6.77 | 6.91 |
| $D_{mean\text{-}heart}$ (%) | 1.72 | 0.53 | 0.98 | 1.66 |
| $D_{mean\text{-}eso}$ (%) | 1.73 | 1.77 | 1.76 | 1.73 |
| $D_{mean\text{-}cord}$ (%) | 3.52 | 0.00 | 1.60 | 1.73 |
| $D_{mean\text{-}body}$ (%) | 3.58 | 3.56 | 3.48 | 3.56 |
| ELST(s) | 165.0 | 167.3 | 163.0 | 164.5 |
| SSPT(s) | 48.8 | 18.8 | 31.3 | 38.3 |
| SSWT(s) | 10.0 | 5.07 | 7.46 | 9.09 |
| Total | 223.8 | 191.1 | 201.8 | 211.9 |

### C. Abdomen

For the abdomen case, the total number of energy layers is 68, and we evaluate QC using 20, 30, and 40 energy layers. As shown in TABLE 3, QC with 40 energy layers achieves a plan quality comparable to that of IMPT with 61 energy layers, as evidenced by similar target coverage and OAR sparing. The reduction in the number of energy layers also shortens the delivery time from 191.1 to 166.2 seconds. When the number of energy layers is further reduced to 20 or 30, QC achieves improved dose sparing for the large bowel and body; however, this comes at the cost of compromised target coverage, with the maximum target dose exceeding clinical acceptance thresholds 120%. The dose distribution and DVH plots in Fig. 3 further validates these results.

TABLE 3. Parameters for abdomen. The plan-quality dosimetric quantities from top to bottom: number of used energies (Energy), optimization objective value $f$, conformity index (CI), max target dose $D_{max}$, mean small bowel dose ($D_{mean\text{-}s\text{-}bowel}$), mean large bowel dose ($D_{mean\text{-}l\text{-}bowel}$), mean body dose ($D_{mean\text{-}body}$), energy layer switching time (ELST), spots spill time (SSPT), spots switching time (SSWT), total delivery time (Total). The unit of time is in second.

| Paras \ Model | IMPT | QC-20 | QC-30 | QC-40 |
|---|---|---|---|---|
| Energy | 61 | 20 | 30 | 40 |
| $f$ | 6.40 | 11.1 | 8.37 | 7.39 |
| CI | 0.91 | 0.88 | 0.91 | 0.91 |
| $D_{max}$ (%) | 116.1 | 124.1 | 120.3 | 118.2 |
| $D_{mean\text{-}s\text{-}bowel}$ (%) | 16.6 | 20.2 | 17.6 | 17.0 |
| $D_{mean\text{-}l\text{-}bowel}$ (%) | 0.65 | 0.41 | 0.61 | 0.63 |
| $D_{mean\text{-}body}$ (%) | 4.41 | 3.60 | 3.96 | 4.22 |
| ELST(s) | 117.4 | 96.7 | 101.7 | 109.9 |
| SSPT(s) | 58.5 | 29.8 | 36.8 | 43.8 |
| SSWT(s) | 15.2 | 6.60 | 10.0 | 12.5 |
| Total | 191.1 | 133.1 | 148.5 | 166.2 |

TABLE 4. Robust optimization for lung. The plan-quality dosimetric quantities from top to bottom: number of used energies (Energy), optimization objective value $f$, conformity index (CI), max target dose $D_{max}$, mean lung dose ($D_{mean\text{-}lung}$), mean heart dose ($D_{mean\text{-}heart}$), mean esophagus dose ($D_{mean\text{-}eso}$), mean spin cord dose ($D_{mean\text{-}cord}$), mean body dose ($D_{mean\text{-}body}$), energy layer switching time (ELST), spots spill time (SSPT),

| Paras \ Model | IMPT | QC-20 | QC-30 | QC-40 |
|---|---|---|---|---|
| Energy | 55 | 20 | 30 | 40 |
| $f$ | 3.46 | 5.97 | 4.00 | 3.59 |
| CI | 0.89 | 0.80 | 0.87 | 0.89 |

spots switching time (SSWT), total delivery time (Total). The unit of time is in second.

| Model \ Paras | IMPT | QC |
|---|---|---|
| Energy | 56 | 35 |
| $f$ | 3.81 | 4.07 |
| CI | 0.81 | 0.80 |
| $D_{max}$ (%) | 115.1 | 116.8 |
| $D_{mean-lung}$ (%) | 7.56 | 7.49 |
| $D_{mean-heart}$ (%) | 2.23 | 2.28 |
| $D_{mean-eso}$ (%) | 2.30 | 1.48 |
| $D_{mean-cord}$ (%) | 2.99 | 2.07 |
| $D_{mean-body}$ (%) | 3.86 | 3.81 |
| ELST(s) | 173.9 | 173.4 |
| SSPT(s) | 49.5 | 34.8 |
| SSWT(s) | 8.61 | 7.32 |
| Total | 232.0 | 215.4 |

*D. Robust optimization*

In this subsection, we assess the robustness of the proposed method. We present results for the lung case, while results for the HN and abdomen cases are provided in the supplementary material Section B. The practical algorithm described in subsection 2.3 (Fig. 1) is employed to adaptively select the number of energy layers, $N_E$. The parameter ε is set to 0.1 to ensure that the plan quality between IMPT and QC remains within an acceptable range.

As shown in TABLE 4, QC required 21 fewer energy layers (56 vs. 35) while maintaining a comparable plan quality to IMPT. For instance, in terms of target coverage, the CI and maximum target dose were 0.75 and 116.1 in IMPT, compared to 0.74 and 117.3 in QC. Furthermore, QC achieved improved OAR sparing compared to IMPT, with the exception of a minor increase in the mean heart dose (2.23 vs. 2.28), QC reduced the mean lung dose from 7.56 to 7.49, mean esophagus dose from 2.30 to 1.48, mean spinal cord dose from 2.99 to 2.07, and mean body dose from 3.86 to 3.81. Additionally, the reduction in the number of energy layers led to a shorter delivery time, decreasing from 232.0 seconds to 215.4 seconds. The dose distributions in Fig. 5 further confirms these findings.

Moreover, QC demonstrated robustness comparable to that of IMPT. As illustrated in Fig. 5, both visual inspection and quantitative analysis reveal a similar DVH band width, along with comparable variations (with respect to nominal case) in target $D_{95}$% ($\Delta D_{95}$%) and maximum OAR dose ($\Delta D_{max}$). These findings indicate that the reduction in energy layers achieved by QC does not compromise plan robustness.

TABLE 5. Comparison for lung. The plan-quality dosimetric quantities from top to bottom: number of used energies (Energy), optimization objective value $f$, conformity index (CI), max target dose $D_{max}$, mean lung dose ($D_{mean-lung}$), mean heart dose ($D_{mean-heart}$), mean esophagus dose ($D_{mean-eso}$), mean spin cord dose ($D_{mean-cord}$), mean body dose ($D_{mean-body}$), energy layer switching time (ELST), spots spill time (SSPT), spots switching time (SSWT), total delivery time (Total). The unit of time is in second.

| Model \ Paras | IMPT | CARD | CARD2 | QC |
|---|---|---|---|---|
| Energy | 56 | 40 | 35 | 35 |
| $f$ | 3.81 | 4.15 | 4.86 | 4.07 |
| CI | 0.81 | 0.81 | 0.78 | 0.80 |
| $D_{max}$ (%) | 115.1 | 118.0 | 116.8 | 116.7 |
| $D_{mean-lung}$ (%) | 7.56 | 7.54 | 7.62 | 7.49 |
| $D_{mean-heart}$ (%) | 2.23 | 2.27 | 2.35 | 2.28 |
| $D_{mean-eso}$ (%) | 2.30 | 2.01 | 1.81 | 1.48 |
| $D_{mean-cord}$ (%) | 2.99 | 2.65 | 2.41 | 2.07 |
| $D_{mean-body}$ (%) | 3.86 | 3.82 | 3.88 | 3.81 |
| ELST(s) | 173.9 | 173.4 | 176.4 | 173.4 |
| SSPT(s) | 49.5 | 38.3 | 34.8 | 34.8 |
| SSWT(s) | 8.61 | 7.90 | 7.34 | 7.32 |
| Total | 232.0 | 219.6 | 218.5 | 215.5 |

*E. Comparison*

In this section, we compare QC with other ELO methods to evaluate its performance, specifically considering CARD [14]. For both CARD and QC, we follow the workflow outlined in Fig. 1 to adaptively select the $N_E$. As in subsection D, we set the threshold ε = 0.1. As before, we present the results for the lung case and leave the HN and abdomen cases to the supplementary material.

As observed in TABLE 5, ELO methods effectively reduce the number of energy layers while maintaining plan quality. For example, compared to IMPT, the number of energy layers is reduced from 56 to 40 in CARD and further to 35 in QC. Despite utilizing five fewer energy layers than CARD, QC achieves comparable target coverage, with a maximum target dose of 118.0 v.s. 116.7 and a CI of 0.81 v.s. 0.80, respectively. Moreover, QC demonstrates improved sparing of OARs, including reductions in the mean lung dose from 7.54 to 7.49, mean esophagus dose from 2.01 to 1.48, and mean spinal cord dose from 2.65 to 2.07. We also conduct experiment on CARD with the same number of energy layer as QC (CARD2). It is observed from results that QC remains competitive. The dose and DVH plots in Fig. 6 further validate the results.

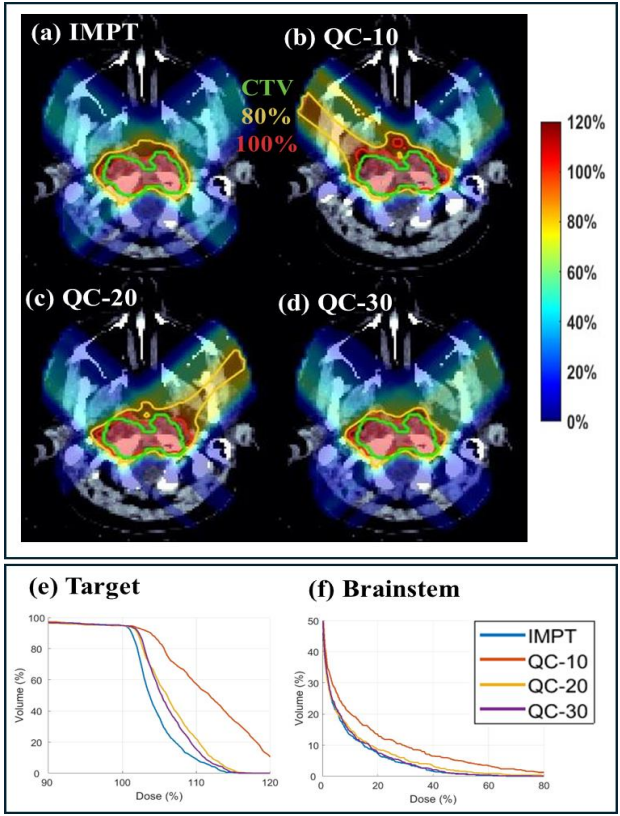

**Fig. 2.** HN. (a) Dose plots of IMPT; (b)-(d) Dose plots of QC with 10, 20 and 30 energy layers, respectively; (e)-(f). DVH plots of target and brainstem. The dose plot window is [0%, 120%] of the prescription dose, with 80% (yellow) and 100% (red) isodose lines and CTV (green) highlighted in dose plots.

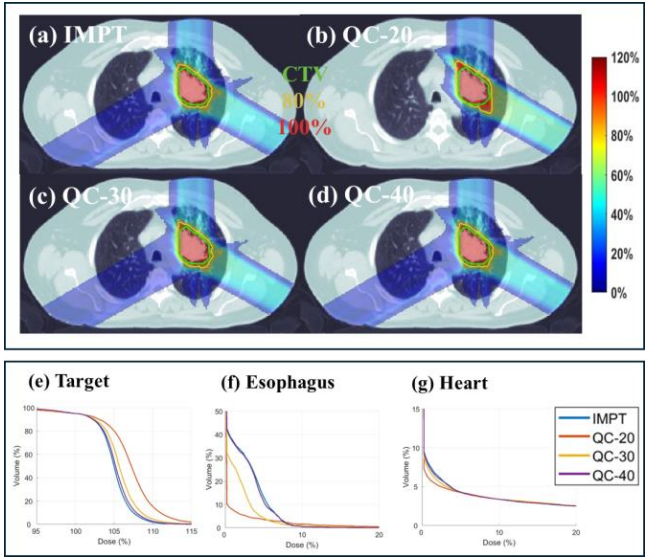

**Fig. 3.** Lung. (a) Dose plots of IMPT; (b)-(d) Dose plots of QC with 20, 30 and 40 energy layers, respectively; (e)-(g). DVH plots of target, esophagus and heart. The dose plot window is [0%, 120%] of the prescription dose, with 80% (yellow) and 100% (red) isodose lines and CTV (green) highlighted in dose plots.

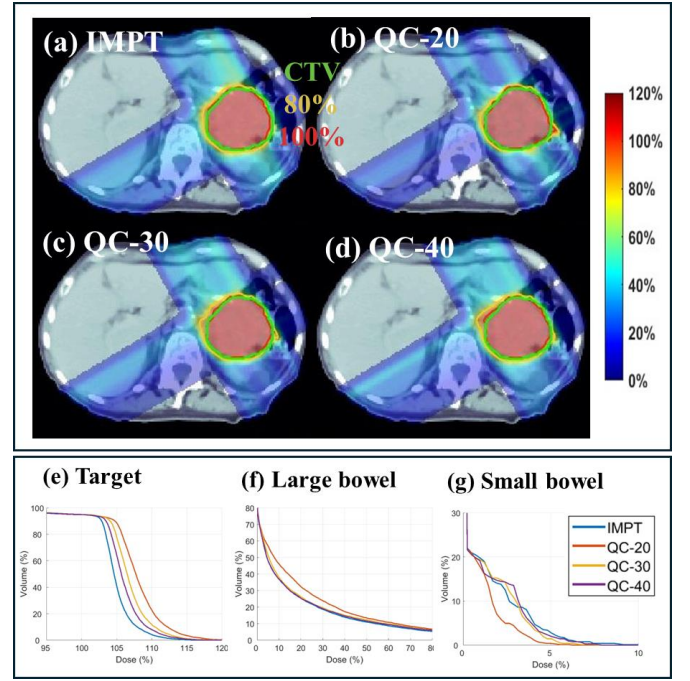

**Fig. 4.** Abdomen. (a) Dose plots of IMPT; (b)-(d) Dose plots of QC with energy layer 20, 30 and 40, respectively; (e)-(g). DVH plots of target, larger bowel and small bowel. The dose plot window is [0%, 120%] of the prescription dose, with 80% (yellow) and 100% (red) isodose lines and CTV (green) highlighted in dose plots.

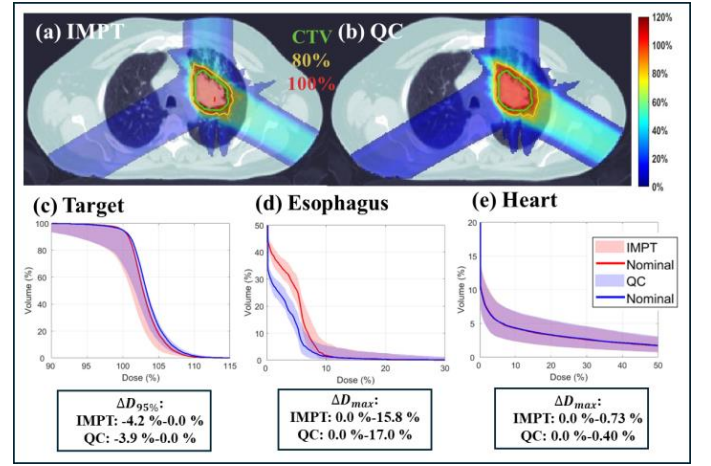

**Fig. 5.** Robust optimization (lung). (a) Dose plots of QC (nominal); (b) Dose plots of QC (nominal); (c)-(e). DVH plots of target, esophagus, and heart. $\Delta D_{95\%}$: variation of $D_{95\%}$ with respect to nominal case for the target. $\Delta D_{max}$: variation of max dose with respect to nominal case for the OAR under consideration. The dose plot window is [0%, 120%] of the prescription dose, with 80% (yellow) and 100% (red) isodose lines and CTV (green) highlighted in dose plots.

## IV. DISCUSSION

In this work, we propose an ELO method based on a mixed-integer model. The model is built upon the ADMM framework, which enables the decoupling of the optimization process into continuous and binary variable subproblems. The subproblem associated with the continuous variables can be efficiently solved using numerical methods such as the conjugate gradient algorithm, which only involves matrix–vector multiplications. The binary variable subproblem admits a quadratic objective form, i.e., a classical QUBO problem, which can be effectively addressed using VQC algorithms. Numerical experiments on various clinical cases have been conducted to validate the correctness and effectiveness of the proposed method. To the best of our knowledge, this is the first work in radiation therapy optimization that solves a mixed-integer problem by integrating first-order methods with variational quantum computing algorithm.

It should be noted that although VQC is employed to solve the QUBO subproblem, the dimensionality of this subproblem remains modest. Specifically, its size corresponds to the number of energy layers, which typically ranges from 50 to 80. This scale is not considered particularly large from the perspective of mixed-integer optimization. Nonetheless, we emphasize that the proposed framework is not limited to the ELO problem and can be extended to other applications in radiation therapy planning. For instance, in beam angle optimization (BAO) [36-37], binary variables can be used to indicate whether a particular delivery angle is selected. Similarly, in the minimum monitor unit (MMU) problem [38-40], binary variables can represent whether a given spot is activated. For these problems, our algorithm can be directly applied, with the only modification being the definition of the binary variable. In these scenarios, the resulting QUBO problem may involve hundreds or even thousands of binary variables, making it significantly more challenging for classical MIP solvers to handle efficiently. In such high-dimensional cases, VQC algorithms may demonstrate a clear computational advantage. While the VQC algorithm in this study is simulated rather than executed on actual quantum hardware, we anticipate its strong potential in future applications. With the increasing availability of commercial quantum computing platforms, such as the IBM Quantum Platform, it becomes particularly appealing to extend our algorithmic framework to larger-scale problems—such as BAO and MMU—on real quantum devices.

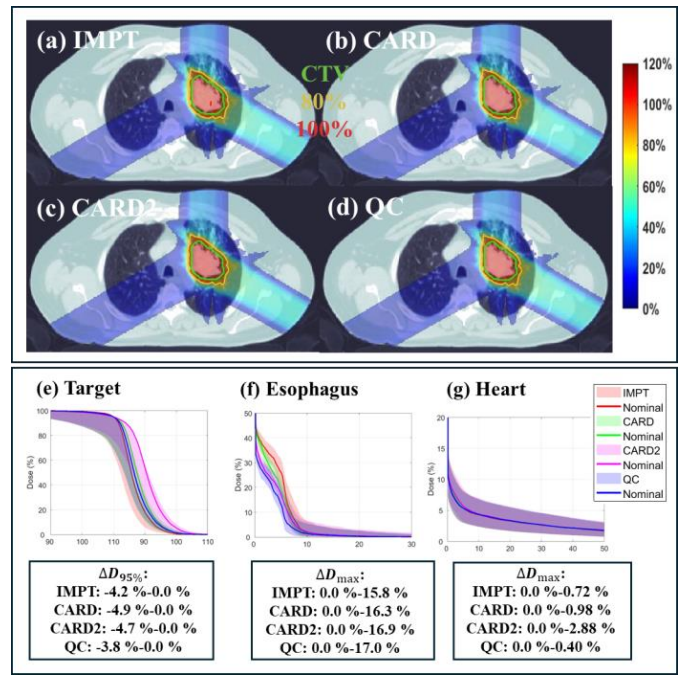

**Fig. 6.** Comparison (**lung**). (a) Dose plots of IMPT (nominal); (b) Dose plots of CARD (nominal); (c) Dose plots of CARD2 (nominal); (d) Dose plots of QC (nominal); (e)-(g). DVH plots of target, esophagus, and heart. $\Delta$D95%: variation of D95% with respect to nominal case for the target. $\Delta$Dmax: variation of max dose with respect to nominal case for the OAR under consideration. The dose plot window is [0%, 120%] of the prescription dose, with 80% (yellow) and 100% (red) isodose lines and CTV (green) highlighted in dose plots.

## V. Conclusion

This work proposed an energy layer optimization method that based on the mixed integer model and variational quantum computing algorithm. Experiments on various site are used to demonstrate the effectiveness and correctness of the method.

## References

[1] A. J. Lomax, T. Bortfeld, G. Goitein, J. Debus, C. Dykstra, P. A. Tercier, P. A. Coucke, and R. O. Mirimanoff, "A treatment planning inter-comparison of proton and intensity modulated photon radiotherapy," *Radiother. Oncol.*, vol. 51, no. 3, pp. 257–271, Dec. 1999, doi: 10.1016/S0167-8140(99)00097-6.

[2] V. Verma, M. V. Mishra, and M. P. Mehta, "A systematic review of the cost and cost-effectiveness studies of proton radiotherapy," *Cancer*, vol. 122, no. 10, pp. 1483–1501, May 2016, doi: 10.1002/cncr.29842.

[3] A. J. Lomax, "Intensity modulated proton therapy and its sensitivity to treatment uncertainties 1: the potential effects of calculational uncertainties," *Phys. Med. Biol.*, vol. 53, no. 4, pp. 1027–1042, Feb. 2008, doi: 10.1088/0031-9155/53/4/014.

[4] A. J. Lomax, "Intensity modulated proton therapy and its sensitivity to treatment uncertainties 2: the potential effects of inter-fraction and inter-field motions," *Phys. Med. Biol.*, vol. 53, no. 4, pp. 1043–1056, Feb. 2008, doi: 10.1088/0031-9155/53/4/015.

[5] K. Ohara, T. Okumura, M. Akisada, T. Inada, T. Mori, H. Yokota, and M. J. Calaguas, "Irradiation synchronized with respiration gate," *Int. J. Radiat. Oncol. Biol. Phys.*, vol. 17, no. 4, pp. 853–857, Aug. 1989, doi: 10.1016/0360-3016(89)90108-5.

[6] S. Mori, T. Inaniwa, T. Furukawa, S. Zenklusen, T. Shirai, and K. Noda, "Effects of a difference in respiratory cycle between treatment planning

and irradiation for phase-controlled rescanning and carbon pencil beam scanning," *Br. J. Radiol.*, vol. 86, no. 1028, p. 20130163, May 2013, doi: 10.1259/bjr.20130163.

[7] J. Dueck, A. C. Knopf, A. Lomax, F. Albertini, G. F. Persson, M. Josipovic, and P. M. af Rosenschöld, "Robustness of the voluntary breath-hold approach for the treatment of peripheral lung tumors using hypofractionated pencil beam scanning proton therapy," *Int. J. Radiat. Oncol. Biol. Phys.*, vol. 95, no. 1, pp. 534–541, Sep. 2016, doi: 10.1016/j.ijrobp.2016.02.022.

[8] M. Hillbrand and D. Georg, "Assessing a set of optimal user interface parameters for intensity-modulated proton therapy planning," *J. Appl. Clin. Med. Phys.*, vol. 11, no. 4, pp. 93–104, Fall 2010, doi: 10.1120/jacmp.v11i4.3362.S.

[9] S. Van de Water, A. C. Kraan, S. Breedveld, W. Schillemans, D. N. Teguh, H. M. Kooy, and M. S. Hoogeman, "Improved efficiency of multi-criteria IMPT treatment planning using iterative resampling of randomly placed pencil beams," *Phys. Med. Biol.*, vol. 58, no. 19, pp. 6969–6983, Oct. 2013, doi: 10.1088/0031-9155/58/19/6969.

[10] S. Van De Water, H. M. Kooy, B. J. Heijmen, and M. S. Hoogeman, "Shortening delivery times of intensity modulated proton therapy by reducing proton energy layers during treatment plan optimization," *Int. J. Radiat. Oncol. Biol. Phys.*, vol. 92, no. 2, pp. 460–468, Oct. 2015, doi: 10.1016/j.ijrobp.2015.02.015.

[11] W. Cao *et al.*, "Proton energy optimization and reduction for intensity-modulated proton therapy," *Phys. Med. Biol.*, vol. 59, no. 21, pp. 6341–6356, Nov. 2014, doi: 10.1088/0031-9155/59/21/6341.

[12] M. F. Jensen, L. Hoffmann, J. B. Petersen, D. S. Møller, and M. Alber, "Energy layer optimization strategies for intensity-modulated proton therapy of lung cancer patients," *Med. Phys.*, vol. 45, no. 10, pp. 4355–4363, Oct. 2018, doi: 10.1002/mp.13138.

[13] Y. Lin, B. Clasie, T. Liu, M. McDonald, K. M. Langen, and H. Gao, "Minimum-MU and sparse-energy-layer (MMSEL) constrained inverse optimization method for efficiently deliverable PBS plans," *Phys. Med. Biol.*, vol. 64, no. 20, p. 205001, Oct. 2019, doi: 10.1088/1361-6560/ab3e6c.

[14] B. Lin, Y. Li, B. Liu, S. Fu, Y. Lin, and H. Gao, "Cardinality-constrained plan-quality and delivery-time optimization method for proton therapy," *Med. Phys.*, vol. 51, no. 7, pp. 4567–4580, Jul. 2024, doi: 10.1002/mp.15789.

[15] A. Wang, Y. N. Zhu, J. Setianegara, Y. Lin, P. Xiao, Q. Xie, and H. Gao, "Development and experimental validation of an in-house treatment planning system with greedy energy layer optimization for fast IMPT," *arXiv preprint*, arXiv:2411.18074, Nov. 2024. [Online]. Available: https://arxiv.org/abs/2411.18074

[16] X. R. Zhu, N. Sahoo, X. Zhang, D. Robertson, H. Li, S. Choi, A. K. Lee, and M. T. Gillin, "Intensity modulated proton therapy treatment planning using single-field optimization: the impact of monitor unit constraints on plan quality," *Med. Phys.*, vol. 37, no. 3, pp. 1210–1219, Mar. 2010, doi: 10.1118/1.3301576.

[17] Y. Lin, H. Kooy, D. Craft, N. Depauw, J. Flanz, and B. Clasie, "A greedy reassignment algorithm for the PBS minimum monitor unit constraint," *Phys. Med. Biol.*, vol. 61, no. 12, pp. 4665–4677, Jun. 2016, doi: 10.1088/0031-9155/61/12/4665.

[18] H. Gao, B. Clasie, T. Liu, and Y. Lin, "Minimum MU optimization (MMO): an inverse optimization approach for the PBS minimum MU constraint," *Phys. Med. Biol.*, vol. 64, no. 12, p. 125022, Jun. 2019, doi: 10.1088/1361-6560/ab23a2.

[19] T. Bortfeld, "Clinically relevant intensity modulation optimization using physical criteria," in *Proc. XII Int. Conf. Use Comput. Radiat. Therapy*, Med. Phys. Publ., 1997, pp. 45–48.

[20] Q. Wu and R. Mohan, "Algorithms and functionality of an intensity modulated radiotherapy optimization system," *Med. Phys.*, vol. 27, no. 4, pp. 701–711, Apr. 2000, doi: 10.1118/1.598914.

[21] H. Gao, "Hybrid proton-photon inverse optimization with uniformity-regularized proton and photon target dose," *Phys. Med. Biol.*, vol. 64, no. 10, p. 105003, May 2019, doi: 10.1088/1361-6560/ab1a3e.

[22] H. Gao, B. Lin, Y. Lin, S. Fu, K. Langen, T. Liu, and J. Bradley, "Simultaneous dose and dose rate optimization (SDDRO) for FLASH proton therapy," *Med. Phys.*, vol. 47, no. 12, pp. 6388–6395, Dec. 2020, doi: 10.1002/mp.14557.

[23] W. Li, Y. Lin, H. Li, R. Rotondo, and H. Gao, "An iterative convex relaxation method for proton LET optimization," *Phys. Med. Biol.*, vol. 68, no. 5, p. 055002, Mar. 2023, doi: 10.1088/1361-6560/acb5f7.

[24] W. Li, W. Zhang, Y. Lin, *et al.*, "Fraction optimization for hybrid proton-photon treatment planning," *Med. Phys.*, vol. 50, no. 12, pp. 6789–6801, Dec. 2023, doi: 10.1002/mp.16636.

[25] H. Gao, J. Liu, Y. Lin, *et al.*, "Simultaneous dose and dose rate optimization (SDDRO) of the FLASH effect for pencil-beam-scanning proton therapy," *Med. Phys.*, vol. 49, no. 3, pp. 2014–2025, Mar. 2022, doi: 10.1002/mp.15418.

[26] Gabay D, Mercier B., 1976. A dual algorithm for the solution of nonlinear variational problems via finite element approximation. Computers & mathematics with applications, 2(1): 17-40.

[27] Glowinski R, Marroco A., 1975. On the approximation by finite elements of order one, and resolution, penalisation-duality for a class of nonlinear Dirichlet problems[J]. ESAIM: Mathematical Modelling and Numerical Analysis-Mathematical Modelling and Numerical Analysis, 9(R2): 41-76.

[28] Boyd, S., Parikh, N., Chu, E., Peleato, B., & Eckstein, J., 2011. Distributed optimization and statistical learning via the alternating direction method of multipliers. Foundations and Trends® in Machine learning, 3(1), 1-122.

[29] Goldstein T, Osher S., 2009. The split Bregman method for L1-regularized problems. SIAM journal on imaging sciences, 2(2): 323-343.

[30] Gao, H., 2016. Robust fluence map optimization via alternating direction method of multipliers with empirical parameter optimization. Physics in Medicine & Biology, 61(7), 2838.

[31] M. R. Hestenes and E. Stiefel, "Methods of conjugate gradients for solving linear systems," *NBS J. Res.*, vol. 49, no. 1, pp. 409–436, 1952.

[32] C. Grange, M. Poss, and E. Bourreau, "An introduction to variational quantum algorithms for combinatorial optimization problems," *4OR*, vol. 21, no. 3, pp. 363–403, Sep. 2023, doi: 10.1007/s10288-023-00534-0.

[33] H. P. Wieser, E. Cisternas, N. Wahl, S. Ulrich, A. Stadler, H. Mescher, and M. Bangert, "Development of the open-source dose calculation and optimization toolkit matRad," *Med. Phys.*, vol. 44, no. 6, pp. 2556–2568, Jun. 2017, doi: 10.1002/mp.12242.

[34] L. Zhao, G. Liu, S. Chen, J. Shen, W. Zheng, A. Qin, and X. Ding, "Developing an accurate model of spot-scanning treatment delivery time and sequence for a compact superconducting synchrocyclotron proton therapy system," *Radiat. Oncol.*, vol. 17, no. 1, p. 87, May 2022, doi: 10.1186/s13014-022-02039-0.

[35] L. Zhao, G. Liu, K. Souris, S. Wuyckens, G. Janssens, K. Poels, and X. Ding, "Machine-specific delivery sequence model of compact superconducting synchrocyclotron proton therapy systems – A multi-institutional investigation," *Int. J. Radiat. Oncol. Biol. Phys.*, vol. 114, no. 3, p. e541, Mar. 2022, doi: 10.1016/j.ijrobp.2022.01.546.

[36] R. Kaderka, K. C. Liu, L. Liu, R. VanderStraeten, T. Liu, K. M. Lee, and C. Chang, "Toward automatic beam angle selection for pencil-beam scanning proton liver treatments: A deep learning–based approach," *Med. Phys.*, vol. 49, no. 7, pp. 4293–4304, Jul. 2022, doi: 10.1002/mp.15717.

[37] H. Shen, G. Zhang, Y. Lin, R. L. Rotondo, Y. Long, and H. Gao, "Beam angle optimization for proton therapy via group-sparsity based angle generation method," *Med. Phys.*, vol. 50, no. 6, pp. 3258–3273, Jun. 2023, doi: 10.1002/mp.16104.

[38] H. Gao, B. Clasie, T. Liu, *et al.*, "Minimum MU optimization (MMO): an inverse optimization approach for the PBS minimum MU constraint," *Phys. Med. Biol.*, vol. 64, no. 12, p. 125022, Jun. 2019, doi: 10.1088/1361-6560/ab23a2.

[39] J. F. Cai, R. C. Chen, J. Fan, *et al.*, "Minimum-monitor-unit optimization via a stochastic coordinate descent method," *Phys. Med. Biol.*, vol. 67, no. 1, p. 015009, Jan. 2022, doi: 10.1088/1361-6560/ac465a.

[40] Y. N. Zhu, X. Zhang, Y. Lin, *et al.*, "An orthogonal matching pursuit optimization method for solving minimum-monitor-unit problems: Applications to proton IMPT, ARC and FLASH," *Med. Phys.*, vol. 50, no. 8, pp. 4710–4720, Aug. 2023, doi: 10.1002/mp.16242.

> REPLACE THIS LINE WITH YOUR MANUSCRIPT ID NUMBER (DOUBLE-CLICK HERE TO EDIT) <